\documentclass[a4paper,12pt]{article}
\usepackage{epsfig}
\usepackage{amsmath}
\usepackage{amssymb}
\usepackage{epsf}
\usepackage{graphicx}
\usepackage{cite}
\usepackage{graphics}
\usepackage{ulem}
\usepackage{rotating}

\newdimen\nude\newbox\chek
\def\slash#1{\setbox\chek=\hbox{$#1$}\nude=\wd\chek#1{\kern-\nude/}}

\newcommand{\sqrtspa}{\sqrt{s_{_{p {\rm A}}}}}

\newcommand{\jpsi}{J/\psi}
\newcommand{\sqrtspsin}{\sqrt{s_{_{\jpsi {\rm N}}}}}
\newcommand{\mpsi}{m_{_{\jpsi}}}

\newcommand{\A}{{\rm A}}

\newcommand{\TA}{T_{_\A}}
\newcommand{\dd}{{\rm d}\,}
\newcommand{\ndf}{{\rm ndf}}
\newcommand{\sig}{\sigma_{_{J/\psi {\rm N}}}}

\def\cO#1{{{\cal{O}}}\left(#1\right)}
\def\pt{p_{_\perp}}
\def\xf{x_{\rm F}}
\def\X{{\rm X}}
\newcommand{\rpa}{R_{_{\A/p}}}
\newcommand{\rwbe}{R_{_{{\rm W}/{\rm Be}}}}
\newcommand{\rag}{R^{^\A}_{_G}}

\begin{document}

\setcounter{footnote}{3}
\renewcommand{\thefootnote}{\fnsymbol{footnote}} 	

\begin{flushright}
LAPTH-1244/08
\end{flushright}

%%%%%%%%%%%%%%%%%%%%%%%%%%%%%%%%%%%%%%%%%%%%%%%
\begin{center}
{\Large\bf Constraints on nuclear gluon densities\\[0.5cm] from $\jpsi$ data}
\end{center}

\begin{center}
{\large  Fran\c{c}ois Arleo\footnote{Email address: \texttt{arleo@lapp.in2p3.fr}}}\\[0.5cm]
{\it LAPTH\footnote{Laboratoire d'Annecy-le-Vieux de Physique Th\'eorique, UMR5108}, Universit\'e de Savoie, CNRS,\\ BP 110, 74941 Annecy-le-Vieux cedex, France}
\end{center}
%%%%%%%%%%%%%%%%%%%%%%%%%%%%%%%%%%%%%%%%%%%%%%%

\begin{abstract}
The gluon density in nuclei, $G^{\rm A}(x)$, is poorly constrained at all $x$ from the current DIS and Drell-Yan data. In this paper, we point out that $\jpsi$ production measured in proton-nucleus collisions at $\sqrtspa=38.8$~GeV by the E866 collaboration puts stringent constraints on the $x$-dependence of the $G^{\rm W}/G^{\rm Be}$ ratio in the $x=2\ 10^{-2}$--$10^{-1}$ range. The E866 data suggest a rather mild $x$-dependence  of $G^{\rm W}/G^{\rm Be}$, and consequently tend to favour DS and HKM sets, rather than the DSg, EKS, and EPS parametrizations which exhibit a large shadowing and anti-shadowing.
\end{abstract}

\setcounter{footnote}{0}
\renewcommand{\thefootnote}{\arabic{footnote}}

A lot of effort has been devoted over the past ten years in order to constrain the parton distributions in nuclei. The first set of nuclear parton densities (nPDFs) at leading order (LO) was given by Eskola, Kolhinen, Ruuskanen and Salgado (EKS)~\cite{Eskola:1998iyEskola:1998df}. This parametrization used deep inelastic scattering (DIS) data to determine the valence-quark densities, and Drell-Yan production in proton-nucleus collisions to constrain the sea. As first done in~\cite{Gousset:1996xt}, the gluon sector in EKS is probed indirectly through the $Q^2$-dependence of the structure function ratios measured by the NMC collaboration. However, the rather large error bars of those data could not allow for a precise determination of the gluon nuclear distributions. The EKS set has been widely used ever since to predict the nuclear dependence of a large variety of hard observables: Drell-Yan, electroweak bosons, large-$\pt$ hadrons, heavy quark and heavy-quarkonium, jets, and prompt photons. A few years later, Hirai, Kumano, Miyama and Nagai (HKM) proposed an alternative distribution, based on an explicit $\chi^2$ analysis of the data~\cite{Hirai:2001npHirai:2004wq} The kinematic dependence of the EKS and HKM sets is shown to be rather different, as discussed e.g. in~\cite{Armesto:2006ph}, reflecting the relative lack of constraints given by the available data. The first global fit analysis of nPDFs at next-to-leading order (NLO) accuracy has then been performed by De Florian and Sassot (DS)~\cite{deFlorian:2003qf}. Finally, an analysis taking into account non-linear corrections in the QCD evolution at small $x$ is also proposed in~\cite{Eskola:2002yc}.

In order to further constrain the gluon nuclear density --~whose knowledge is essential to perform reliable predictions in proton-nucleus and nucleus-nucleus collisions at the LHC~--, Eskola, Paukkanen, and Salgado (EPS) recently included in their analysis~\cite{Eskola:2008ca} the forward hadron production data measured in $d$--Au collisions by the BRAHMS collaboration at RHIC~\cite{Arsene:2004ux}. The rather large suppression in $d$--Au with respect to $p$--$p$ collisions seen experimentally leads to a dramatic effect in the  ratio $\rag(x)\equiv G^{\rm A}(x)/G^p(x)\ll 1$ at small $x=\cO{10^{-3}}$ and, following momentum sum rules, to a significant anti-shadowing, $\rag(x)\sim 1.4$ at $x=\cO{10^{-1}}$ and at rather low scales. As a consequence, the $x$-dependence of $G^{\rm A}/G^p(x)$ in EPS proves much steeper than the one predicted in any other nPDF sets.

In this paper, we investigate $\jpsi$ production in proton-nucleus collisions as an interesting channel to constrain the gluon nuclear density. Indeed, assuming that it follows that of open charm, $\jpsi$ production proceeds {\it via} gluon fusion and quark-antiquark annihilation. However, as long as $\jpsi$ is produced at small longitudinal momentum fraction, $\xf\ll 1$, the gluon fusion channel dominates over the $q\bar{q}$ annihilation process. At leading order, the differential cross section is therefore simply proportional to the product of gluon densities:
\begin{equation}
  \label{eq:cem}
  \frac{\dd \sigma}{\dd{x_1}\dd{x_{_2}}}(p \ \A \to \jpsi\ \X) \propto G^{p}(x_{_1}, Q^2) \ G^{\A}(x_{_2}, Q^2)\ \delta(x_1 x_{_2} s - \mpsi^2),
  \end{equation}
where 
\begin{equation}\label{eq:x1x2}
  x_{_{1, 2}} = \frac{1}{2} \, \left( \sqrt{x_{_{\rm F}}^2+4 \ \mpsi^2/s} \pm x_{_{\rm F}} \right)
\end{equation}
are the projectile and target-parton momentum-fractions ($\sqrt{s}$ being the centre-of-mass energy of the hadronic collision), and the typical scale $Q$ at which the gluon densities should be evaluated is given by the $\jpsi$ mass scale, $Q=\cO{\mpsi}$. In this kinematics, the nuclear production ratio,
\begin{equation}
  \rpa(x_{_2}) = \frac{1}{A}\ \
  \frac{\dd\sigma}{\dd{x_{_2}}}(p+\A\to\jpsi+\X)
  \Big/ \frac{\dd\sigma}{\dd{x_{_2}}}(p+p\to\jpsi+\X),
\end{equation}
therefore reduces to $\rpa(x_{_2}) \simeq\ \rag(x_{_2})$, which we would like precisely to constrain. This relationship was first used in~\cite{Gupta:1992be} to extract the gluon nuclear density from $\jpsi$ data in $\pi$--A and $p$--A collisions measured by the NA3 and E772 collaborations.

However, $\jpsi$ is a pretty weakly bound state which may be sensitive to inelastic rescattering processes in large nuclei, which will spoil the above simple relationship between $\rpa$ and $\rag$. Assuming the factorization between the $c\bar{c}$ production and the rescattering process, the nuclear production ratio can be written as:
\begin{equation}
  \label{eq:ratioapprox}
  \rpa(x_{_2}) \simeq\ \rag(x_{_2})\ \times\ S_{\rm{abs}}(\A, \sig),
  \end{equation}
where $S_{\rm{abs}}(\A, \sig)$ denotes the probability for no interaction (or ``survival probability'') of the $\jpsi$ meson with the target nucleus. It depends of course on both the atomic mass number $A$ of the nucleus and the $\jpsi$--N inelastic cross section, $\sig$, and is given in a Glauber model by\cite{Capella:1988ha}
\begin{equation}
  \label{eq:supp}
  S_{\rm{abs}}(\A, \sig) = \frac{1}{(A-1) \ \sig}\, \int \dd {\bf b} \left( 1 - e^{- (1-1/A) \ \TA({\bf b}) \ \sig} \right),
\end{equation}
where $\TA({\bf b})$ is the so-called nuclear thickness function.
Because of the present lack of constraints on $\sig$ (and therefore on $S_{\rm{abs}}$), the hope to determine $\rag(x)$ from $\jpsi$ suppression data in $p$--A collisions is rather limited. It is the reason why this channel has never been considered in the global fit analyses of nPDFs. As we shall see later, significant constraints can nevertheless be achieved by fitting the 
$\jpsi$ suppression data, letting the $\jpsi$ inelastic cross section as a free parameter.

The present analysis relies in particular on the measurements performed by the E866 fixed-target experiment in $p$--Be and $p$--W collisions at $\sqrtspa=38.8$~GeV. The E866 collaboration reported on the nuclear production ratio~\cite{Leitch:1999ea}:
\begin{equation}
\label{eq:ratio}
  \rwbe(\xf) = \frac{A_{_{\rm Be}}}{A_{_{\rm W}}} \ \
  \frac{\dd\sigma}{\dd\xf}(p+{\rm W}\to\jpsi+\X)
  \Big/ \frac{\dd\sigma}{\dd\xf}(p+{\rm Be}\to\jpsi+\X)
\end{equation}
where $A_{_{\rm Be}}=9$ and $A_{_{\rm W}}=183$ denote the atomic mass numbers. The $\xf$ range covered by these data is fairly large, $-0.1 \lesssim \xf \le 0.9$. At large $\xf$, $\rwbe$ decreases very rapidly down to 0.3 in the largest $\xf$ bin. The reason for such a large suppression is not settled yet. In any case, the $\jpsi$ inelastic interaction together with the modifications of parton densities in nuclei --~the only effects assumed in the present work~-- could explain only {\it partly} this large-$\xf$ suppression~\cite{Vogt:1999dw}. Furthermore, the lack of $x_{_2}$-scaling when comparing large-$\xf$ E866 data points with NA3~\cite{Badier:1983dg} and PHENIX~\cite{Adler:2005ph} data indicates that other effects, beyond inelastic rescattering and nPDFs corrections, are responsible for the suppression reported at large $\xf$. Therefore, the present analysis will be restricted to $\jpsi$ production around mid-rapidity, $|\xf| \le 0.25$, which corresponds to the range $x_{_2}=2\ 10^{-2}$--$10^{-1}$ in the target momentum-fraction.

Even though Eq.~(\ref{eq:ratioapprox}) should be a good approximation for the $\jpsi$ suppression in proton-nucleus collisions, the calculation is performed in the colour evaporation model\footnote{This model describes heavy-quarkonium production near threshold and reproduces fairly well rapidity and low-$\pt$ spectra in hadronic collisions, although it is known to face some difficulties (see e.g.~\cite{Hoyer:1998haLansberg:2006dh} for a critical discussion). In the present context, this approach is only used to estimate the small corrections to Eq.~(\ref{eq:ratioapprox}) because of the quark annihilation process; hence our results do not depend much on which specific model is assumed.}, including both the $gg$ and $q\bar{q}$ channels for the production of $c\bar{c}$ pairs with invariant masses $\mpsi \le m \le 2 m_D$ (see Ref.~\cite{Arleo:2006qk} for more detail on the present calculation). The W/Be nuclear production ratio measured by E866 has been fitted using the various nPDFs currently available, DS~\cite{deFlorian:2003qf}, DSg~\cite{deFlorian:2003qf}, EKS~\cite{Eskola:1998iyEskola:1998df}, EPS~\cite{Eskola:2008ca}, HKM~\cite{Hirai:2001npHirai:2004wq}, keeping $\sig$ as a free parameter\footnote{For the time being $\sig$ is taken to be independent of $x_{_2}$, although we shall discuss in the following the implications of a possible $x_{_2}$-variation of the cross section on our results.}. It is not the goal to discuss the actual value of $\sig$ obtained, we refer the reader to~\cite{Arleo:2006qk} for a comprehensive discussion on this point. Rather, the respective agreement between the fits and the E866 measurements using each nPDF set is investigated. Moreover, because of the above mentioned nuclear absorption which strength is not precisely determined, these data will rather probe the $x_{_2}$-dependence of $\rag(x_{_2})$ rather than its absolute value.

\begin{figure}[h]
\begin{minipage}[t]{7.5cm}
  \begin{center}
    \includegraphics[height=7.5cm]{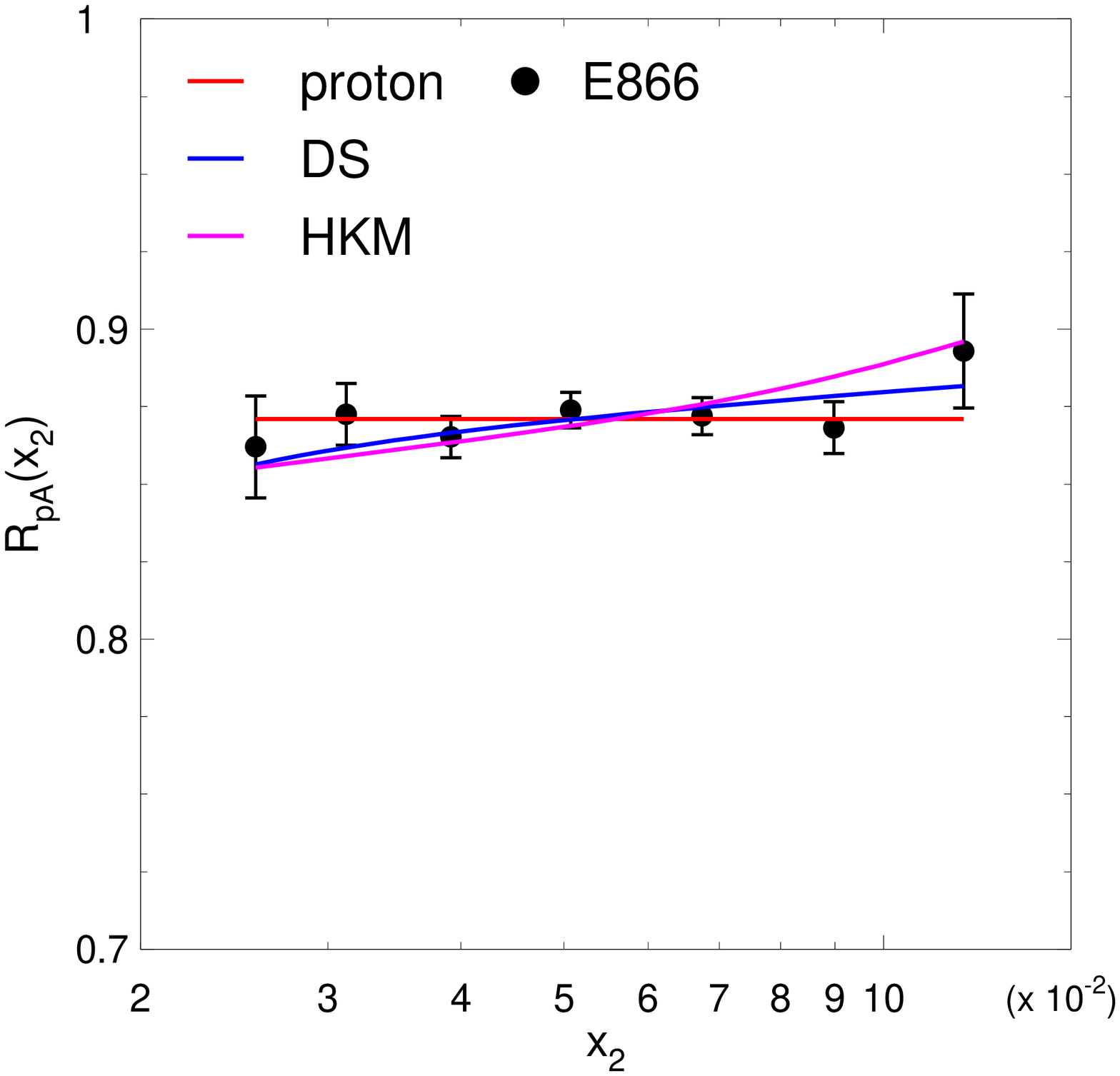}
  \end{center}
\end{minipage}
~\hfill
\begin{minipage}[t]{7.5cm}
  \begin{center}
    \includegraphics[height=7.5cm]{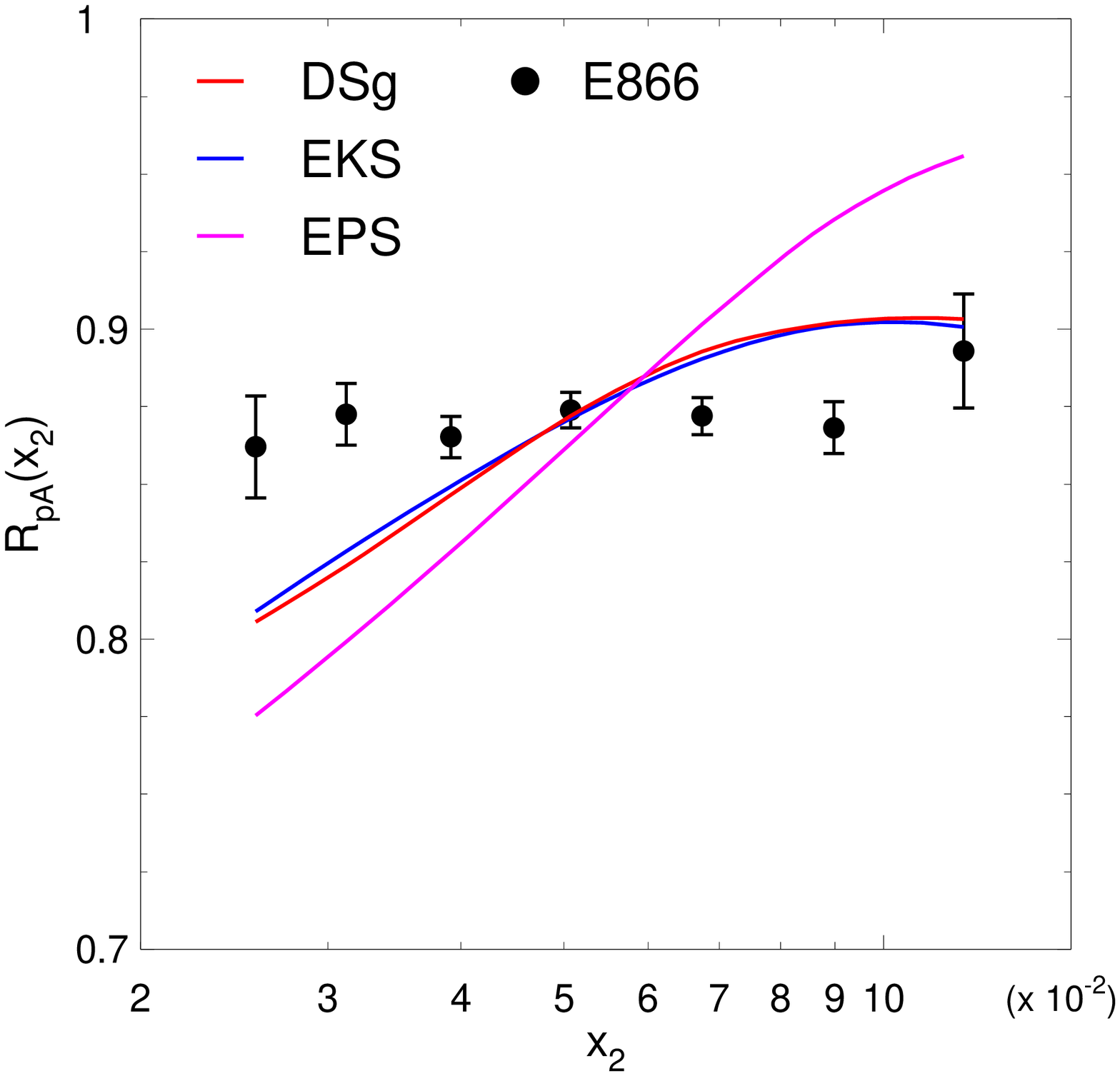}
  \end{center}
\end{minipage}
\caption{The W-over-Be $\jpsi$ production ratio as a function of $x_{_2}$ measured by the E866 collaboration, in comparison with the best fits using the proton, DS, HKM (left) and DSg, EKS, EPS (right) parton densities. The overall $3\%$ systematic errors (i.e. independent of $x_{_2}$) are not shown.}
  \label{fig:e866}
\end{figure}

As can be seen in Fig.~\ref{fig:e866}, the E866 data points show a remarkably flat behaviour as a function of $x_{_2}$. Therefore, following Eq.~(\ref{eq:ratioapprox}), the gluon nuclear density ratio $\rag(x)$ at the $\jpsi$ mass scale seems to have a pretty limited variation, say $\lesssim5\%$ from $x=2\ 10^{-2}$ to $x=10^{-1}$. Let us compare this observation with the predictions given by the different nPDF sets. The trend reported by the E866 data is well reproduced by the DS and HKM sets on the entire $x_{_2}$-domain. These two parametrizations indeed do not predict a strong anti-shadowing at $x_{_2}=\cO{10^{-1}}$ nor a large shadowing at small $x_{_2}$. Moreover, note that the E866 measurements are also consistent with no modifications of the gluon distributions in nuclei, $\rag(x)=1$ (labelled {\it proton}), or at least with an $x$-independent ratio, $\rag(x)=\rag$, because of the uncertainty of the normalization due to the nuclear absorption. The good agreement between the data and these theoretical calculations is also indicated in Table~\ref{tab:chi2} where the $\chi^2/{\rm ndf}$ for each calculation is given and turns out to be $\cO{1}$. On the contrary, the nPDFs sets which predict a strong dependence\footnote{More precisely a much stronger dependence of $R^{\rm W}_{_G}$ with respect to that of $R^{\rm Be}_{_G}$.} of $\rag$ on $x$, like DSg, EKS, and EPS, seem to be somehow disfavoured by these measurements. In the fits shown in Fig.~\ref{fig:e866} (right), the large (EKS, DSg) and very large (EPS) variation assumed in those sets turns out to be in striking contradiction with the data, as also indicated by the large values $\chi^2/{\rm ndf}$ (respectively $10.1$, $12.0$, and $35.9$) given in Table~\ref{tab:chi2}.  It is remarkable that the strong shadowing present in EPS in order to reproduce the large-rapidity hadron production data measured by BRAHMS at RHIC appears to be too large to reproduce the $\jpsi$ data in the $x_{_2}=10^{-2}$--$10^{-1}$ range. It may indicate that it is not possible to fit simultaneously the forward RHIC results and lower energy (larger $x$) data in terms of a universal modification of parton densities in nuclei.

We investigated as well the constraints given by the NA3 measurements in $p$--A collisions at lower beam energy, $\sqrtspa=19.4$~GeV~\cite{Badier:1983dg}. However, the limited $x_{_2}\simeq 7\ 10^{-2}$--$10^{-1}$ coverage (restricting again to not too large $\xf$) as well as the larger error bars do not allow for differentiating the various nPDFs, all predictions leading to $\chi^2/\ndf=\cO{1}$.

\begin{table}[htb]
  \centering
  \begin{tabular}[c]{ccccccc}
 \hline
 \hline
nPDF set & proton  & DS & DSg & EKS & EPS & HKM \\
 \hline
 $\chi^2/\ndf$ &  0.5 &  0.6 &  12.0 &  10.1 &  35.9 & 1.2\\
  \hline
  \hline
  \end{tabular}
  \caption{$\chi^2$/ndf values between the best fits and the E866 data using the proton, DS, DSg, EKS, EPS, and HKM nuclear parton densities.}
  \label{tab:chi2}
\end{table}

As already discussed, one should however keep in mind that the nuclear absorption is an important process at work in the $\jpsi$ suppression. In particular, one could argue that a fast variation of the nuclear absorption as a function of $x_{_2}$ may actually compensate the apparently too strong $x$-dependence of $\rag(x)$ assumed in DSg, EKS, and EPS. In perturbative QCD, $\sig$ depends slightly on the incident $\jpsi$--nucleon center of mass energy, $\sig\propto\left( \sqrt{s_{_{{\jpsi N}}}} \right)^{2\lambda}$ ,  where the exponent $\lambda\simeq0.25$ is related to the behaviour of the small-$x$ gluon distribution in the nucleon, $xG(x)\sim x^{-\lambda}$~\cite{Bhanot:1979vb}. Since  $\sqrtspsin \simeq \mpsi/\sqrt{x_{_2}}$,
  the cross section is expected to scale like $x_{_2}^{-\lambda}$. Expanding the exponential in~(\ref{eq:supp}) leads to $S_{\rm abs}(x_{_2})\simeq 1 - \#/x_{_2}^{\lambda}$ which {\it increases} with $x_{_2}$. The energy dependence of $\sig$ would actually worsen the agreement between the fits and the data. The only possibility to reconcile the rapid $x$-dependence of $\rag$ with the flat behaviour of the E866 data would be to appeal to dramatic formation time effects: when $x_{_2}$ is large, the $\jpsi$ state is formed rapidly (recall that the Lorentz boost $\gamma\propto1/{x_{_2}}$) and could experience a stronger interaction with the nucleus than a more compact $c\bar{c}$, at small $x_{_2}$, propagating through the nuclear medium. The competition between such formation time effects together with the strong $x$-dependence of $\rag(x)$ may then result into an accidentally flat ratio $\rwbe$. This explanation would be really fortuitous and appears therefore rather unlikely. We do not really see either how NLO corrections could alter the present conclusions. Because of the extra gluon radiation, the gluon densities  could be probed at slightly larger values of $x_{_2}$ than what is estimated in Eq.~(\ref{eq:x1x2}) at LO. This would nevertheless not improve the agreement between the data and the DSg, EKS, and EPS fits.
 
 To summarize, we discuss in this paper how $\jpsi$ suppression measurements performed in proton-nucleus collisions ($\sqrtspa=38.8$~GeV) by the E866 experiment~\cite{Leitch:1999ea} gives interesting constraints on the $x$-dependence of the gluon distribution in nuclei. The flat behaviour observed experimentally suggests a small variation of the $G^{\rm A}/G^p(x)$ ratio in the $x=10^{-2}$--$10^{-1}$ range at the $\jpsi$ mass scale, which is consistent with DS and HKM parametrizations. On the contrary, the large shadowing and antishadowing effects (and from this the fast variation of $G^{\rm A}/G^p(x)$) assumed e.g. in DSg, EKS, and EPS seem to be disfavoured, unless fortuitous effects due to the interaction of dynamical $c\bar{c}$ pair in the nuclear medium are present. In that sense, comparing the present E866 $\jpsi$ data with the isolated prompt photon measurements in $d$--Au collisions soon to be taken at RHIC in the $\pt=2$--$5$~GeV range (probing the gluon nPDF at roughly to $x_{_2}=4\ 10^{-2}$--$10^{-1}$~\cite{Arleo:2007js}) will be particularly interesting.

\section*{Acknowledgements}

I would like to thank Patrick Aurenche,Thierry Gousset, and St\'ephane Peign\'e for useful comments on the manuscript.
 
\providecommand{\href}[2]{#2}\begingroup\raggedright\endgroup

\end{document}